\RequirePackage{fix-cm}
\documentclass{svjour3}                     
\smartqed  
\usepackage{graphicx}
\usepackage{gensymb}
\usepackage{amsmath}
\usepackage{amssymb}
\usepackage{aas_macros}
\usepackage[dvipsnames]{xcolor} 

\begin{document}

\title{Cometary science with CUBES\thanks{The datasets analysed during the current study are available from the corresponding author on reasonable request.}
}


\author{Cyrielle Opitom         \and
        Colin Snodgrass         \and
        Fiorangela La Forgia \and
        Chris Evans \and
        Pamela Cambianica   \and
        Gabriele Cremonese \and
        Alan Fitzsimmons \and
        Monica Lazzarin \and
        Alessandra Migliorini 
}

\institute{C. Opitom, C. Snodgrass\at
              Institute for Astronomy, University of Edinburgh, Royal Observatory, Edinburgh EH9 3HJ, UK \\
              \email{copi@roe.ac.uk}           
           \and
           F. La Forgia, M. Lazzarin\at
           Dipartimento di Fisica e Astronomia G.Galilei, Università di Padova, Vicolo dell'Osservatorio 3, 35122 Padua, Italy
           \and
           C. J. Evans\at
           UKATC, Royal Observatory, Blackford Hill, Edinburgh, EH9 3HJ, UK 
           \and
           P. Cambianica, G. Cremonese \at
           INAF Astronomical observatory of Padova, Vicolo dell’Osservatorio 5, 35122 Padova, Italy
           \and
           A. Fitzsimmons\at
           Astrophysics Research Centre, School of Mathematics and Physics, Queens University Belfast, Belfast BT7 1NN, UK 
           \and
           A. Migliorini\at
           Institute for Space Astrophysics and Planetology, IAPS-INAF, Rome, Italy
}


%
%


\date{Received: date / Accepted: date}

\maketitle

\begin{abstract}
The proposed CUBES spectrograph for ESO's Very Large Telescope will be an exceptionally powerful instrument for the study of comets. The gas coma of a comet contains a large number of emission features in the near-UV range covered by CUBES (305-400 nm), which are diagnostic of the composition of the ices in its nucleus and the chemistry in the coma. Production rates and relative ratios between different species reveal how much ice is present and inform models of the conditions in the early solar system. In particular, CUBES will lead to advances in detection of water from very faint comets, revealing how much ice may be hidden in the main asteroid belt, and in measuring isotopic and molecular composition ratios in a much wider range of comets than currently possible, provide constraints on their formation temperatures. CUBES will also be sensitive to emissions from gaseous metals (e.g., FeI and NiI), which have recently been identified in comets and offer an entirely new area of investigation to understand these enigmatic objects.
\end{abstract}

\section{Introduction}
\label{intro}

Comets are key bodies to study the formation and evolution of the solar system. They were formed about 4.6 billion years ago and have remained mainly unprocessed ever since. Cometary ices thus retain invaluable clues about the conditions prevailing in the early solar system. Comets are also of great interest from the astrobiology point of view as potential vectors to deliver water to the early Earth. Comets are now stored in two main reservoirs: the Kuiper Belt beyond the orbit of Neptune, and the Oort Cloud at the very edge of the solar system. A third reservoir of comets was also identified recently in the main asteroid belt, between the orbits of Mars and Jupiter.

\begin{figure*}
\centering
\includegraphics[width=11cm]{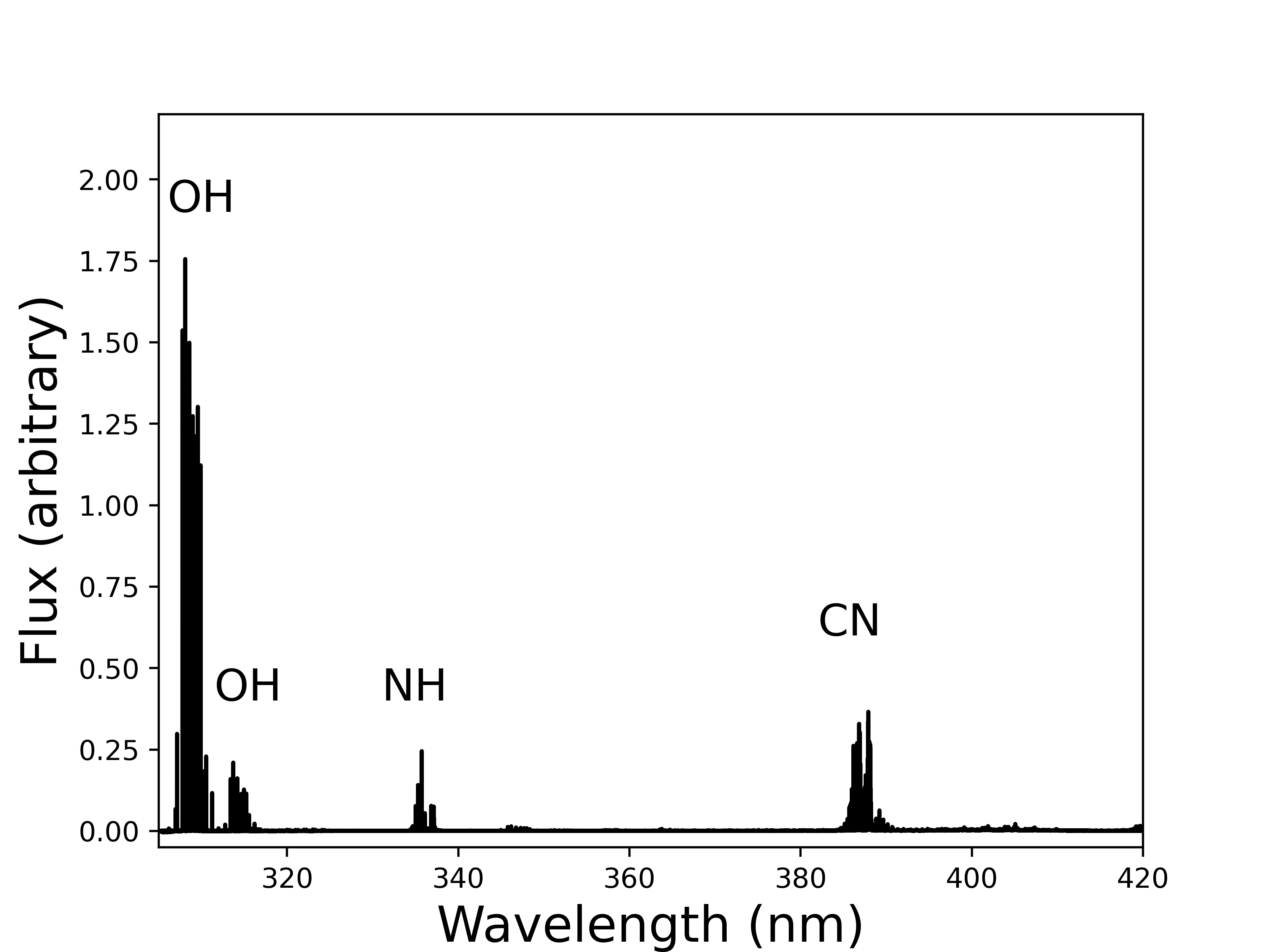}
  \caption{Spectrum of comet C/2002 T7 (LINEAR) with VLT/UVES in the 305-420 nm range showing the main emission bands from the comae of comets that will be covered by CUBES.}
   \label{T7_Broad}
\end{figure*}

The composition of cometary ices is often studied through the observation of the coma, the atmosphere around the comet created by the sublimation of nuclear ices when the comet is sufficiently close to the Sun. Numerous species can be detected through visible spectroscopy of the comae of comets, in particular in the near-UV region. For instance, a typical comet spectrum over the 305-420\,nm range is shown in Fig.~\ref{T7_Broad}, with strong emission bands from OH, NH, and CN. Emissions from several ions (CO$_2^+$, CO$^+$, and N$_2^+$) and metals (FeI and NiI) can also be detected at these wavelengths. 

Improved spectral sensitivity at near-UV wavelengths will enable considerable progress in the study of cometary ices, their composition, and the formation and evolution of planetesimals in the early solar system. To deliver significantly better performance at these short wavelengths the new Cassegrain U-Band Efficient Spectrograph (CUBES) instrument is now in development for the Very Large Telescope (VLT). The CUBES design is optimised for efficiency in the near-UV (305-400\,nm, with a goal of 300-420\,nm) range. This will give a higher transmittance than current near-UV spectrographs, enabling a deeper search for water in the asteroid belt, measurement of the D/H ratio in the comae of comets, measurement of the N$_2$/CO ratio in a number of comets, and more sensitive detection of gaseous iron and nickel in comets. 

The CUBES project completed its Phase~A conceptual design in mid 2021, and will move into the construction phase in 2022; technical details of the design are outlined elsewhere in this volume \cite{Zanutta2022}. A key aspect of the design for cometary observations is the provision of two resolving powers by using two (exchangeable) image slicers, each with six slices \cite{Calcines2022}. These provide a high-resolution (HR) mode with $R$\,$\ge$\,22\,000, with an on-sky field of view of 1.5$''$\,$\times$\,10$''$ (and a slice-width of 0.25$''$), and a low-resolution (LR) mode with $R$\,$\ge$\,5\,000, with a larger field on the sky of 6$''$\,$\times$\,10$''$ (and 1$''$ slices).

In this article we introduce the topics where CUBES will provide exciting new capabilities for cometary science, supported by quantitative performance predictions calculated with simulation tools developed during the Phase~A study \cite{Genoni2022}.

\section{Searching for water in the asteroid belt}
\label{water}

The search for water in our solar system is far from complete and is tremendously difficult to undertake from ground-based facilities given the large water content of Earth's atmosphere. Detecting ice on the surface of distant bodies can be achieved with infrared spectroscopy, but only for the largest/brightest objects. For smaller bodies that are too faint to study directly, and/or objects closer to the Sun where any remaining ice is sub-surface, we must look for outgassing water in the coma.

Water itself can be detected through vibrational bands at infrared (IR) wavelengths but because of the telluric absorption from water in the Earth's atmosphere, most IR transitions can only be observed with space-based telescopes. Some transitions from `hot bands' can be observed using ground-based, high-resolution IR spectrographs such as NIRSPEC on the Keck telescope or CRIRES on the VLT (see e.g. \cite{DelloRusso2004}). Similarly, rotational transitions can be detected at sub-millimeter wavelengths but, due to atmospheric absorption, have mostly been observed with space telescopes. The water content of cometary nuclei can also be estimated by observing water photo-dissociation products: OH, O, and H. The OH (hydroxyl) radical is a widely used proxy of water vapour in the coma. It has a very strong fluorescence A$^{2}\Sigma^{+}$-X$^{2}\Pi$ (0-0) emission band around 308 nm in the near-UV, as can be seen in Fig.~\ref{T7_Broad}. Observations of the 18 cm $\Lambda$-doublet OH at radio wavelengths as well as forbidden oxygen lines [OI] (in particular the doublet around 630 nm) have been used in the past to estimate the water production of comets \cite{Morgenthaler2001}. Observations of forbidden oxygen lines have proved to be sensitive to low levels of water production, but they can also be produced by CO or CO$_2$ photo-dissociation, making it a less secure proxy for water detection, especially for distant targets. Finally, H (Lyman-$\alpha$) emission can be observed in the far-UV but requires the use of space telescopes.

In their review about main belt comets (MBCs) and ice in the solar system, \cite{Snodgrass2017a} discussed the various methods that can be used to detect water around MBCs and compare their efficiencies. They identified the OH band at 308 nm as one of the most promising features to detect water vapour around a MBC. While observations of H (Lyman-$\alpha$) in the far-UV or H$_2$O directly in the IR from space would be more sensitive, they will be substantially more expensive. 

The OH A-X (0-0) band is currently only observable for relatively active comets while near the Sun and Earth due to a lack of sensitivity of existing instruments at these UV wavelengths. This severely limits studies of water production in comets around their orbits and we miss the seasonal effects that the Rosetta mission has recently revealed to be important (e.g. \cite{Kramer2017}). It also prevents study of the vast majority of comets which are simply too faint. Even more tantalising are studies of MBCs – bodies in asteroidal orbits that are seen to undergo activity (usually detected by a dust tail or trail) which is thought to arise from sublimation (e.g. recurrent activity near perihelion). They have typical sizes (km-scale or smaller) that are very common in the asteroid belt, so detection of outgassing water would point to a potentially large population of icy bodies, hence a large reservoir of water, of considerable interest in the context of models of the formation and evolution of the inner solar system \cite{OBrien2018}. Even though sublimation of water ice is the most likely explanation for their activity, water has yet to be detected around a MBC. Confirming the presence of water ice in MBCs would allow us to at least place some constraint on the dust-to-ice ratio of MBCs and to investigate whether they formed in-situ, similarly to some ice-rich asteroids, which would have implications on the position of the snow line and the conditions prevailing in the proto-planetary disk at the time of their formation, or if they formed in the outer solar system and were implanted later on in the asteroid belt, which must be reproduced by solar system dynamical evolution models \cite{Hsieh2014}.

\begin{figure*}
\centering
\includegraphics[width=12cm]{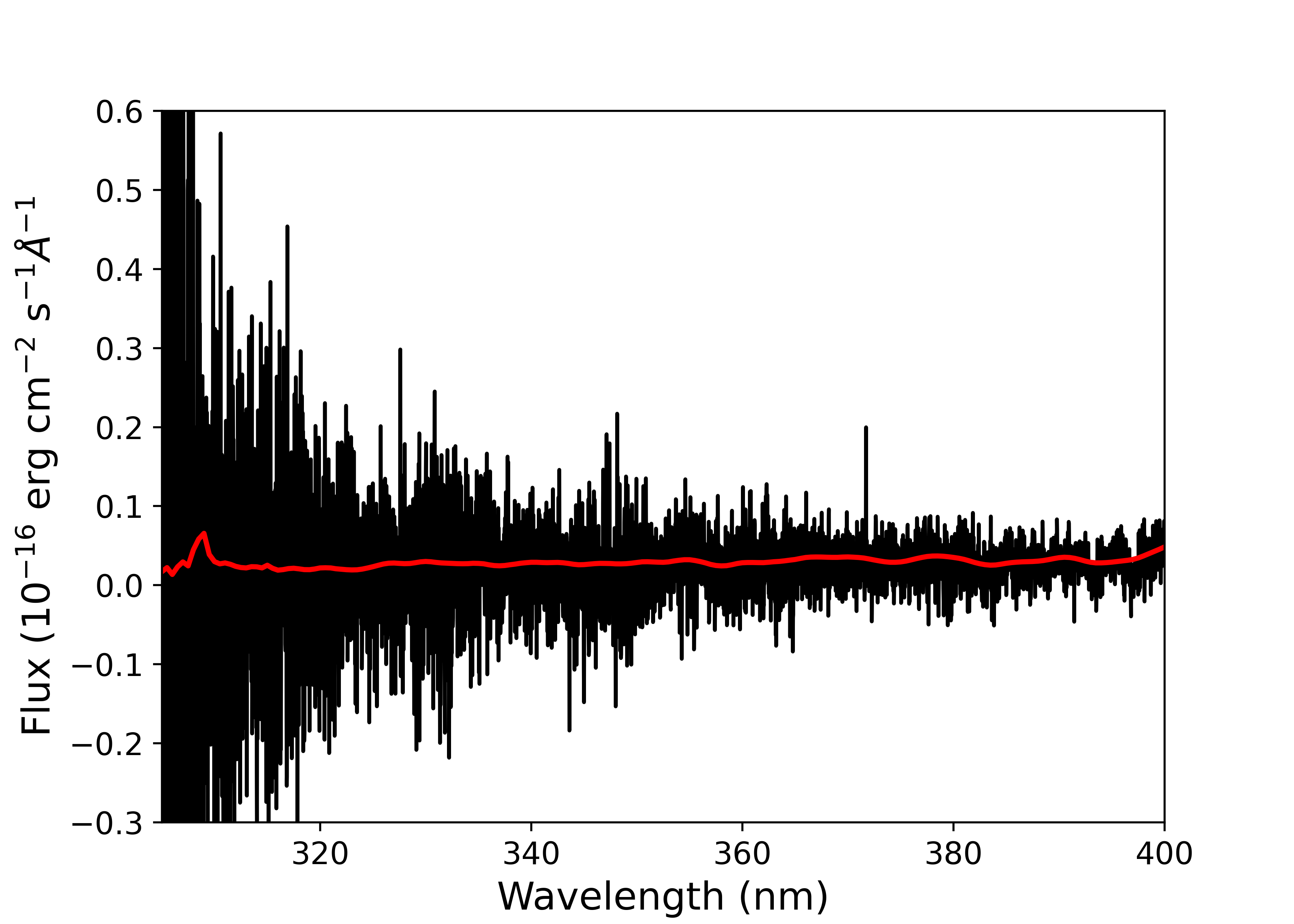}
  \caption{X-Shooter spectrum of main belt comet P/2012 T1 (black) from \cite{Snodgrass2017b}. The red curve is an artificial spectrum of P/2012 T1 produced with the Planetary Spectrum Generator \cite{Villanueva2018} for a water production rate of  $5\times10^{25}$ molecules/s. Significantly better signal-to-noise is needed in the near UV to search for very weak outgassing of OH by main belt comets.}
   \label{2012T1}
\end{figure*}

Observations of the 308 nm OH A-X (0-0) band in MBCs require a high throughput near the atmospheric cut-off. So far, the OH (0-0) band has been observed using either narrow-band filters or spectrographs. Even if narrow-band filters coupled with photometers or CCDs are very efficient to detect OH in faintly active objects, no narrow-band filters are currently installed on 8-m class telescopes. In addition, spectroscopy offers the advantage of making the removal of the underlying dust continuum easier. The observation of a MBC with the VLT/X-Shooter spectrograph shown in Fig. \ref{2012T1} produced only upper limits to OH emission \cite{Snodgrass2017b}, so the throughput of a new UV spectrograph needs to be significantly better at this wavelength than that for the bluest order of X-Shooter. The 2.5h integration of the MBC P/2012 T1 (PANSTARRS) shown in Fig.~2 led to an upper limit of the water production rate of $8\times10^{25}$ molecules/s \cite{Snodgrass2017b}, while estimates of the expected water production rates based on the energy balance or the dust production rate for a typical MBC are between half and two orders of magnitudes lower, in the range 2$\times$10$^{23}$ to 5$\times$10$^{25}$ molecules/s \cite{Snodgrass2017b}. 

The detection of water through the OH (0-0) band at 308 nm does not require high spectroscopic resolution and is one of the cases that motivated the inclusion of the LR mode (with a wider entrance slit) in the CUBES design. Indeed, at heliocentric distances of up to several au comets are extended objects, spanning from a few arcseconds to several arcminutes, or even degrees for the most extended targets. The LR mode, with a larger on-sky field of view will deliver greater sensitivity in the search for water in solar system objects by increasing the amount of light from the source entering the instrument. 


\begin{figure*}
\centering
\includegraphics[width=12cm]{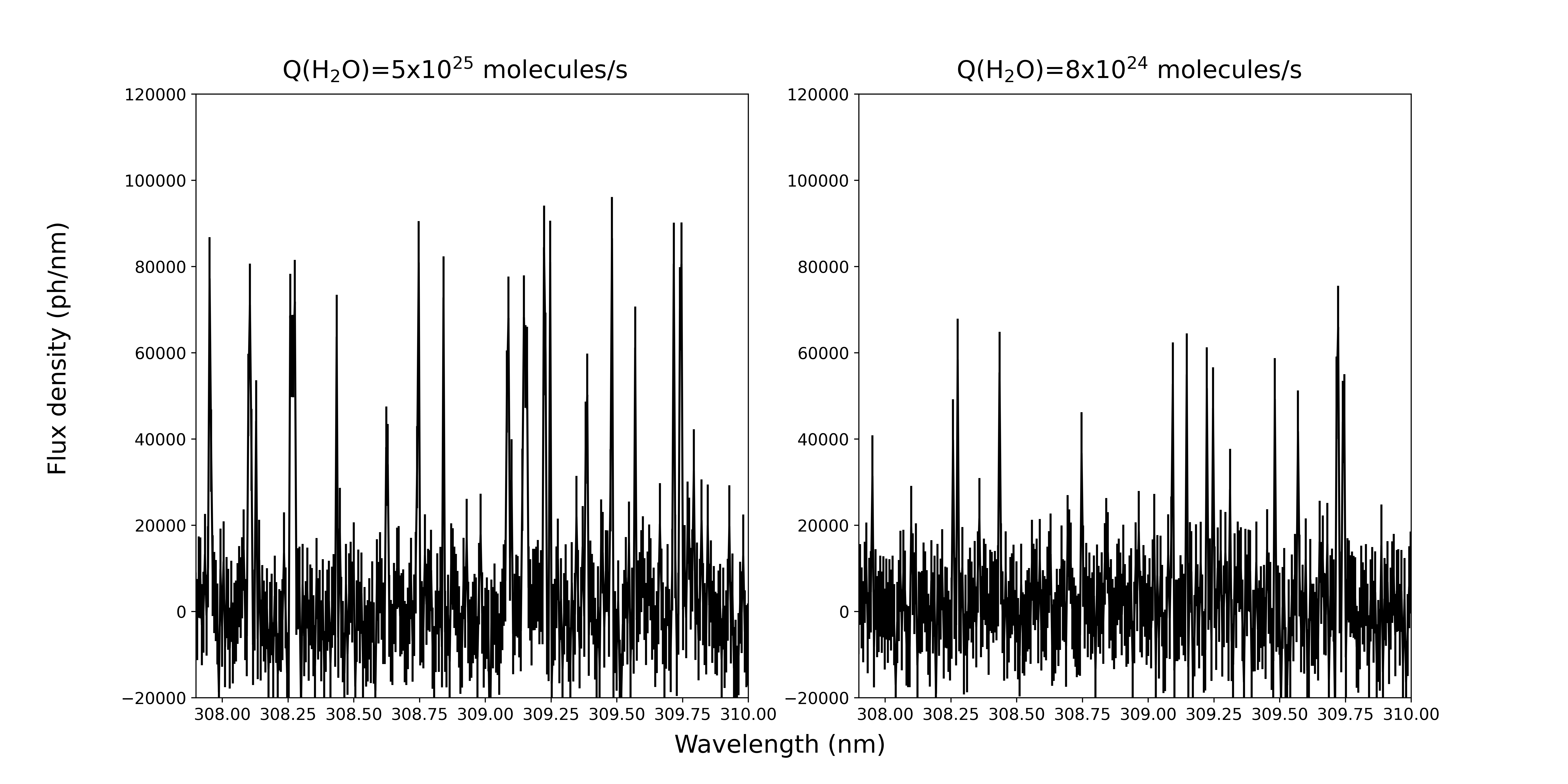}
  \caption{Simulated extracted spectra of main belt comet P/2012 T1 with CUBES around the peak of the OH band for water production rates of $2.5\times10^{25}$ and $8\times10^{24}$ molecules/s. The simulated comet spectra was produced using the Planetary Spectrum Generator \cite{Villanueva2018} combined with the CUBES end-to-end simulator \cite{Genoni2022}. }   \label{Sim2012T1}
\end{figure*}

\medskip
To estimate the performance of CUBES in terms of water detection, we created artificial spectra of MBCs with a range of water production rates ($1\times10^{24}$ to $5\times10^{25}$ molecules/s) using the Planetary Spectrum Generator \cite{Villanueva2018}. We used a dust-to-gas ratio of 1, the typical water lifetime from \cite{Cochran1993}, and realistic observing circumstances for comet 133P. We fed these spectra into the CUBES end-to-end simulator \cite{Genoni2022} for the HR mode and verified for which water production level we could recover a clear detection ($> 3 \sigma$) of individual lines of the OH A-X (0-0) band. From our tests we estimated that we should detect a OH production rate as low as $5\times10^{24}$ molecules/s in a 2h integration on the MBC 133P at perihelion. This is a factor of ten lower than the current upper limits based on attempts to directly detect  water or water dissociation products in the coma of a MBC with the Herschel space observatory or VLT/X-Shooter (see e.g. \cite{Snodgrass2017a}). The estimated water production rate of 133P is between $1.6\times10^{24}$ molecules/s (predicted from the energy balance at the comet) and of the order of $5\times10^{25}$ molecules/s (based on dust production rates) \cite{Snodgrass2017a}. We performed similar simulations for the MBC P/2012 T1, and Fig.~\ref{Sim2012T1} shows the simulated observations for 1h integrations for total water production rates of $2.5\times10^{25}$ and $8\times10^{24}$ molecules/s. This demonstrates that CUBES would be able to detect water at least at a level ten times lower than the upper limit set with X-Shooter.

Depending on the actual water production rate of a given MBC, 2h integrations with CUBES should allow us to detect OH directly or, in the worst-case scenario, to put more definitive upper limits than those currently available, and at least rule out some proposed models for MBC activity.  Simulations with the CUBES exposure time calculator \cite{Genoni2022} show that, even for a moderately extended object of a few arcseconds, using the LR mode would give a gain of more than a factor of two in signal-to-noise at the wavelength of the OH band compared to the HR simulations outlined above, allowing detection of even lower levels of water production.

In addition to detecting water in the main belt, the high sensitivity of CUBES at the wavelength of the OH band presents an opportunity to measure water production rates in distant comets. With current facilities, water is usually only detected out to about 3 au. Cometary activity has been detected out to more than 20 au pre-perihelion (and even more for C/1995 O1 Hale-Bopp post-perihelion), distances at which it is likely driven by ices more volatile than water such as CO or CO$_2$ \cite{Meech2017,Hui2019,Zhang2019}. However, the activity of distant comets and the transition between water-driven activity and activity driven by more volatile ices is still poorly understood. The capacity to detect OH in the comae of comets at larger distances from the Sun enabled by CUBES will provide critical insights into the activity of distant comets. CUBES' timeline and sensitivity will also make it a powerful tool to characterize potential targets of the Comet Interceptor mission \cite{Snodgrass2019}, to be launched in 2029 and whose target will be selected while far away from the Sun. 

\section{Measuring the D/H ratio in the comae of comets}
\label{d/h}

\begin{figure*}
\centering
\includegraphics[width=12cm]{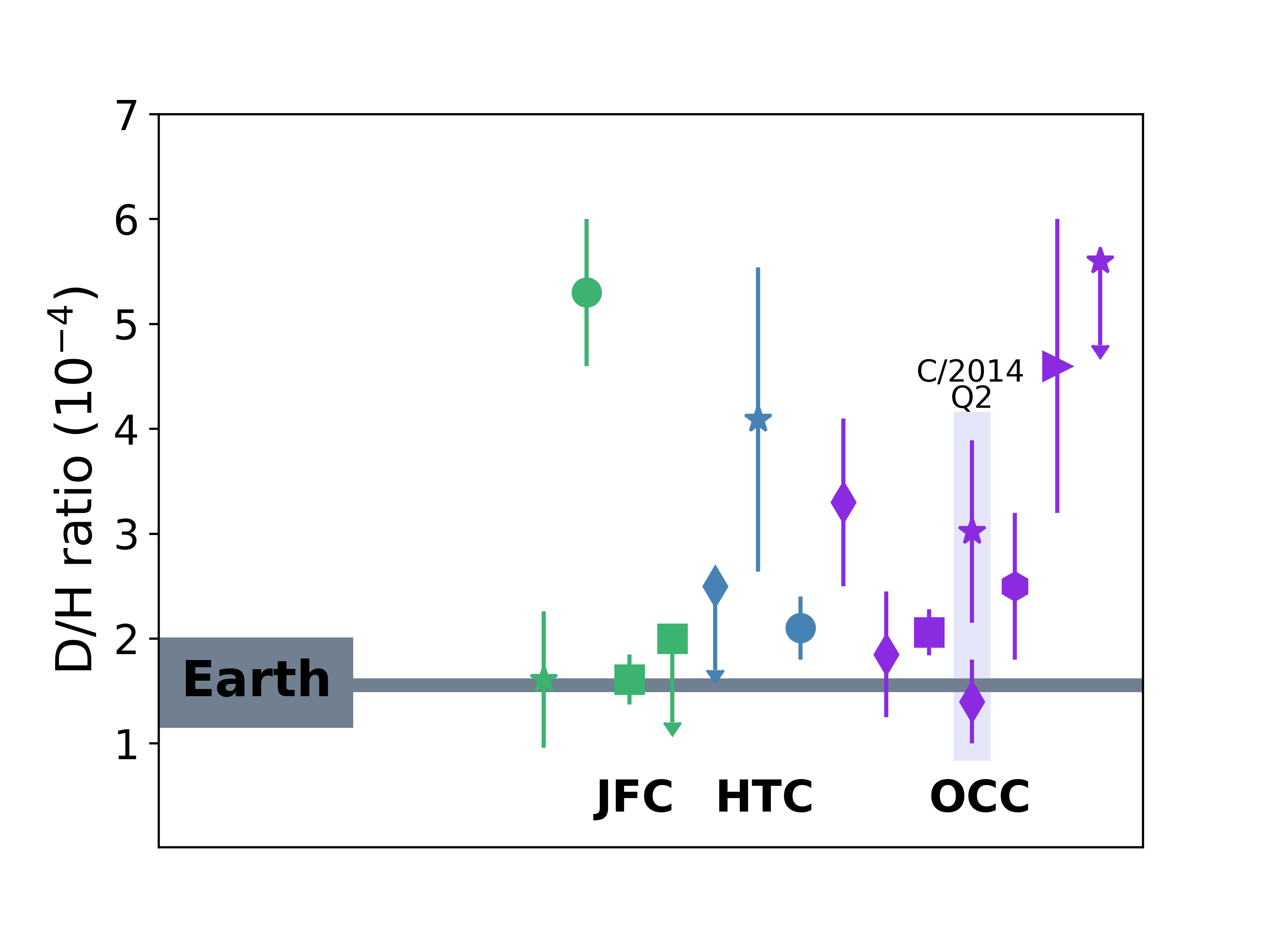}
  \caption{D/H ratios measured in cometary water for different types of comets: JFC (Jupiter Family Comets), HTC (Halley Type Comets), OCC (Oort Cloud Comets). Values from \cite{Bockelee2015,Lis2019,Paganini2017}. Different symbols reflect measurements with different techniques: circles -- {\it in situ} by spacecraft; squares -- far-IR (Herschel space telescope or SOFIA); diamonds -- sub-mm spectroscopy; stars -- high-resolution near-IR spectroscopy; hexagon -- high-resolution UV spectroscopy; triangle -- HST far-UV spectroscopy.}
   \label{DHcomets}
\end{figure*}

Isotopic abundance ratios measured in cometary material are particularly important to decipher conditions prevailing in the early solar system as they are very sensitive to local physico-chemical conditions through fractionation processes \cite{Bockelee2015}. The deuterium to hydrogen (D/H) ratio has a particular importance as it is used to investigate the fate of water in the solar system and the origin of terrestrial water. Measurements of the D/H ratio in cometary water have only been performed so far for just over a dozen comets -- unusually bright ones, or two measured {\it in situ} by spacecraft. The first measurements in long-period comets, originating from the Oort Cloud, revealed values about twice larger than that measured for the Earth’s oceans \cite{Bockelee2015}. In contrast, measurements of the D/H ratio in Jupiter Family comets (coming from the Kuiper Belt) originally revealed a value consistent with the Earth’s oceans \cite{Hartogh2011,Lis2013}. However subsequent measurements, including those performed by the Rosetta mission in the coma of 67P/Churyumov-Gerasimenko, suggested that a similar diversity of the D/H ratio exists for both families of comets \cite{Altwegg2015,Biver2016,Lis2019}. Fig. \ref{DHcomets} shows the diversity of D/H ratio measurements in cometary water. This diversity is currently being interpreted as comets forming over a wide range of distances from the young Sun before being scattered to their current reservoirs, e.g. the Oort Cloud or the Kuiper Belt.

The main problem in interpreting the current measurements of D/H ratios in comets is that most of them have been performed with different instruments, over different wavelength ranges, and focusing on different molecules (H$_2$O or its dissociation products). The symbols in Fig. \ref{DHcomets} reflect measurements made with different techniques. The measurements thus lack homogeneity and might suffer from systematics, due to different instruments/techniques and using different models to derive the D/H ratio. Measurement of the D/H ratio with different techniques for the same comet has only been done for a single comet: C/2014 Q2 (Lovejoy) \cite{Biver2016,Paganini2017}. As can be seen in Fig. \ref{DHcomets}, these two values are different by about a factor two. This illustrates the need to measure the D/H ratio consistently for a sample of comets with a single instrument. This is the only way to allow for a reliable comparison between comets while avoiding potential systematic effects from different techniques.

The D/H ratio in cometary water can be measured from the ground in the near-UV using the bright OH and OD A-X (0-0) and (1-1) bands around 308 and 312 nm. Currently, this has only been done for comet C/2002 T7 (LINEAR), represented by the purple hexagon in Fig.~\ref{DHcomets} \cite{Hutsemekers2008}, because of the lack of sensitivity of near-UV high resolution spectrographs. A high-efficiency spectrograph down to 305 nm will open-up completely new opportunities for the measurement of D/H in comets from ground-based observations of OD/OH. Additionally, a resolving power of $R$\,$\ge$\,20\,000 (corresponding to $\Delta\lambda$\,$\ge$\,0.01 nm) is adequate to identify the lines. Indeed, the OH and OD lines from the A-X (0-0) and (1-1) bands are separated by $\Delta\lambda$\,$\gtrsim$\,1 nm \cite{Hutsemekers2008}. The high spectral resolution of CUBES compared to the resolution necessary to separate the two isotopes will be useful to avoid chance coincidences between OD and other OH lines. By using a fluorescence model to retrieve the OD/OH isotopic ratio, we can then estimate D/H in cometary water (\cite{Hutsemekers2008}). 
With its high sensitivity, CUBES could also allow us to measure the D/H ratio for the first time in an interstellar object like 2I/Borisov, which would provide crucial information about the planetary formation process in its origin system. While photodissociation products of water were detected in the coma of 2I/Borisov \cite{Opitom2021,McKay2020,Xing2020}, it was not active or bright enough for measurements of the D/H ratio with current facilities. 

Simulated spectra created with the Planetary Spectrum Generator \cite{Villanueva2018} for a comet of typical composition and dust-to gas ratio of 1 show that for a comet with a total gas production rate of $3\times10^{28}$ molecules/s and assuming the ratio OD/OH found by \cite{Hutsemekers2008} (4$\times$10$^{-4}$), the strongest OD lines at 307.5 nm \cite{AHearn1985} should have a flux density of the same order of magnitude of a typical OH band from a comet with a water production rate of $5\times10^{24}$ molecules/s, which we have shown above we can detect with CUBES even for a comet in the main belt. Given that the fluorescence efficiency of OD relative to OH strongly depends on the heliocentric velocity of the comet \cite{AHearn1985}, we assumed the most favourable heliocentric velocity for OD detection and this yields a fluorescence efficiency of $1.77\times10^{-15}$ erg/s/molecule for the strongest OD line. In such favourable conditions we should be able to detect OD for water production rates of at least $3\times10^{28}$ molecules/s with about a 2h integration time with CUBES. Currently, D/H ratios can only be measured with UVES for comets with a water production rate of about $10^{30}$ molecules/s, so CUBES will significantly improve our ability to measure its ratio in cometary water.


\section{Measuring the N$_2$/CO ratio in comets}
\label{N2/CO}

\begin{figure*}
\centering
\includegraphics[width=12cm]{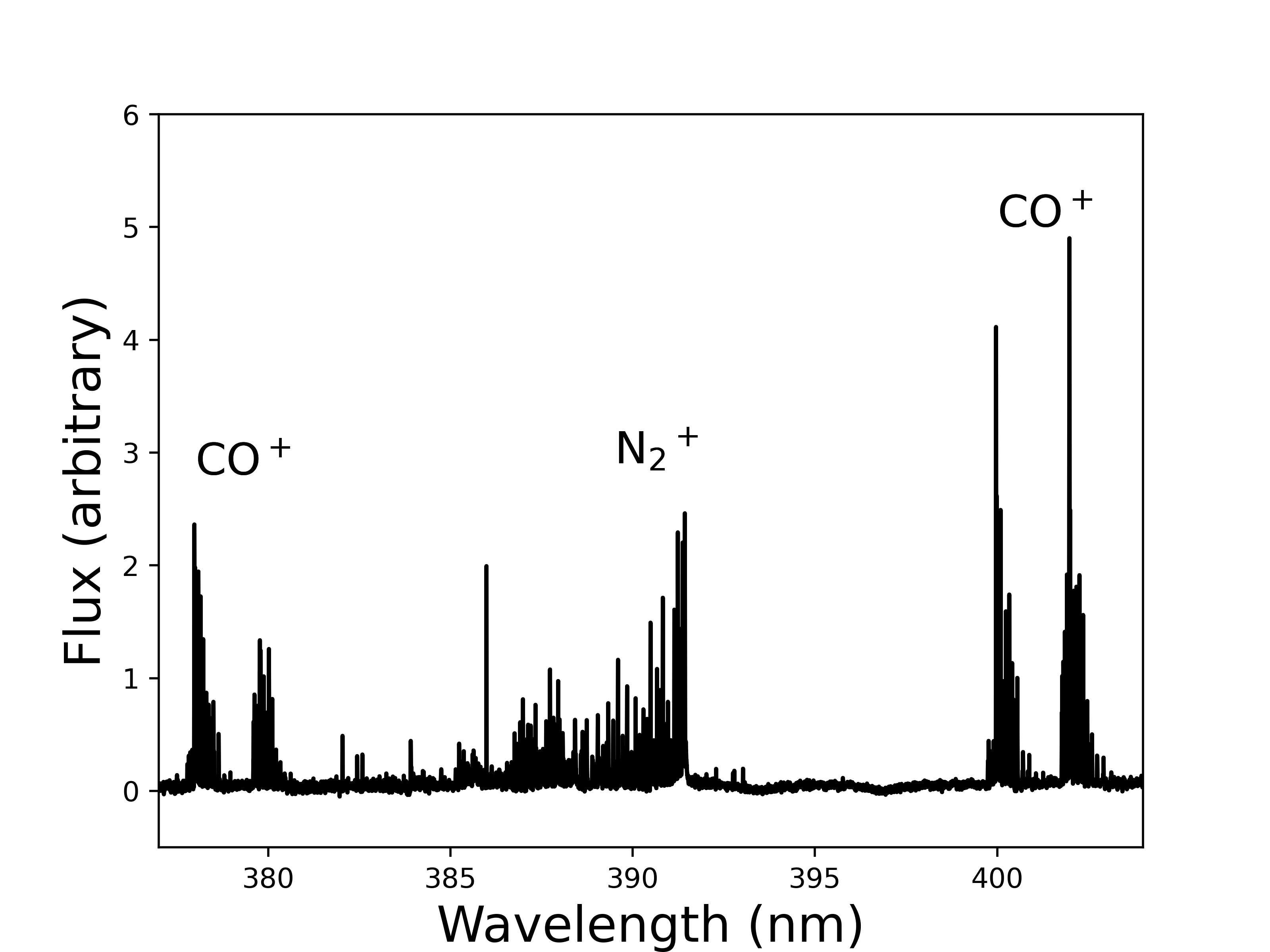}
  \caption{Spectrum of comet C/2016 R2 (PANSTARRS) with VLT/UVES in the 375-405 nm range showing the N$_2^+$ and CO$^+$ emission bands (from \cite{Opitom2019}).}
   \label{R2_ions}
\end{figure*}

The ratio between molecular nitrogen and carbon monoxide abundances (N$_2$/CO) in cometary ices is very sensitive to the formation temperature of comets. Measuring this ratio is therefore important to decipher models of planetesimal formation and constrain the physical properties of the solar nebula at the time of comet formation. This ratio has only been estimated for a handful of comets so far, because N$_2$ is very difficult to detect directly with ground-based observatories. The only direct measurement of the ratio was made for comet 67P/Churyumov-Gerasimenko from mass spectroscopy with the ROSINA instrument onboard Rosetta \cite{Rubin2015,Rubin2019}. However, the N$_2$/CO ratio can also be indirectly estimated by measuring the ratio between the  N$_2^+$ and CO$^+$ ions, produced by the ionisation of N$_2$ and CO in the comae of comets (we note that CO$^+$ can also be produced by the photo-dissociation of CO$_2$). The N$_2^+$ and CO$^+$ ions both have several emission bands in the blue part of the visible spectrum, with the brightest located around 391 nm for N$_2^+$ ($\mathrm{B^2\Sigma_u^+-X^2\Sigma_g^+}$ (0-0) band) and around 378, 400, and 425 nm for the CO$^+$ $\mathrm{A^2\Pi-X^2\Sigma}$ (4,0), (3,0), and (2,0) bands (see Fig. \ref{R2_ions}).

Simultaneous detection of the N$_2^+$ and CO$^+$ emission bands is hard to achieve because the intensity of the emission bands is usually low compared to other species detected in the visible spectrum of comets. Detection of the N$_2^+$ emission band at 391 nm was claimed for a few comets in the 1990s but these detections were contested because of potential confusion with atmospheric N$_2^+$ emission. More reliable detections were only reported recently for comets C/2002 VQ94 (LINEAR), 29P/Schwassmann-Wachmann 1, and C/2016 R2 (PANSTARRS) \cite{Korsun2014,Ivanova2016,Cochran2018}. The reason why N$_2^+$ emission has only been detected in a handful of comets is likely a combination of the low sensitivity of most high-resolution spectrograph in the near-UV and lower N$_2$ abundance or ionisation in some comets. 

A major challenge in trying to detect N$_2^+$ in cometary comae is the difficulty in separating the cometary emission from N$_2^+$ emission coming from the Earth's atmosphere. High spectral resolution observations are required to separate the cometary and telluric components, where $R$\,$\sim$\,20\,000 (i.e. velocity resolution of $\sim$15 km s$^{-1}$) is sufficient for a significant range of geocentric velocities (e.g. \cite{Hughes2000}).

In addition to CO$^+$ and N$_2^+$, the CO$_2^+$ ion also has emission bands
((0,0,0) $\mathrm{\tilde{A}^2\Pi_{u,3/2}} - \mathrm{(0,0,0)\tilde{X}^2\Pi_{g,1/2}}$ and (0,0,0)$\mathrm{\tilde{A}^2\Pi_{u,3/2} - (0,0,0)\tilde{X}^2\Pi_{g,3/2}}$) in the 350-365~nm range. CO$_2^+$ can serve as a proxy to estimate the CO$_2$ abundance in cometary ices, a measurement that is very difficult to perform as CO$_2$ can only be detected directly from space.

Estimating the abundance of neutral species in the coma of comets using ions is not straightforward. Ions have a spatial distribution in the coma very different from neutrals because they are accelerated by the solar wind and this needs to be accounted for to retrieve abundances of neutrals \cite{Jockers1999}. Whether the ions can be used to derive the abundance of neutral species also depends on if the ions are mainly produced through photoionization of the corresponding neutrals or if they are produced through other processes like ion–neutral chemistry. For example, for comet C/2016 R2 (PANSTARRS), it has be shown that the CO$_2^+$ was also produced through charge exchange between CO$^+$ and CO$_2$, indicating that for comets with very high CO abundances, CO$_2^+$ might not be a reliable indicator of the CO$_2$ abundance \cite{Raghuram2021}. 

Nevertheless, in spite of these caveats, ions emission bands observed at optical wavelengths have proven very useful in the past to estimate the relative abundance of difficult to observe neutrals in the coma of comets, and in particular the N$_2$/CO ratio \cite{Korsun2014,Ivanova2016,Cochran2018}. The high sensitivity of CUBES over the 300-400 nm rang will enable estimates of the N$_2$/CO ratio of a sample of comets from different families, providing constraints on their formation conditions. It will also allow us to perform the same measurement on interstellar comets that were formed in very different environments.


\section{Metals in the comae of comets}
\label{metals}

\begin{figure*}
\centering
\includegraphics[width=10cm]{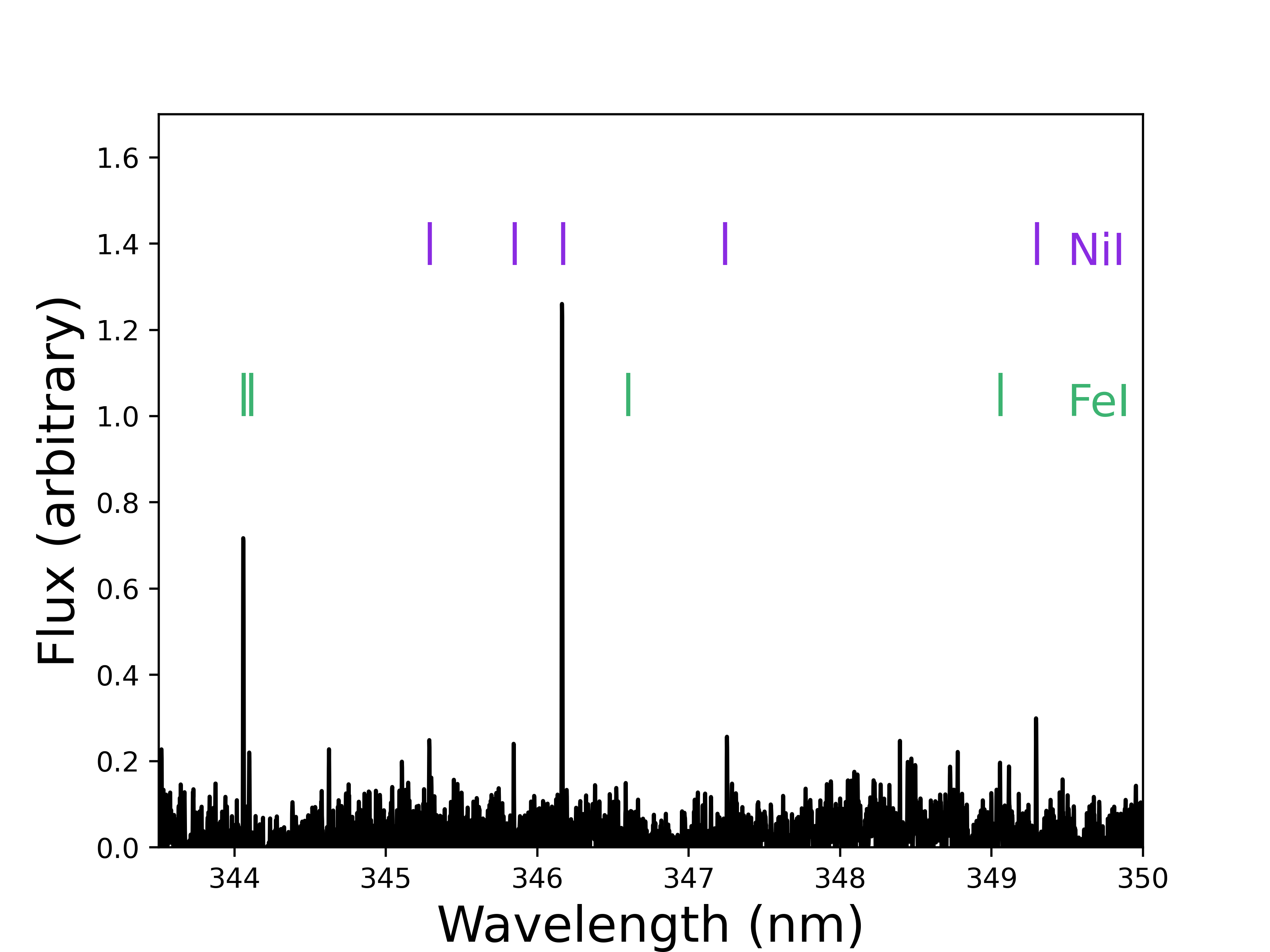}
  \caption{Spectrum of comet C/2016 R2 (PANSTARRS) with VLT/UVES in the 334-350 nm range showing several FeI and NiI emission lines.}
   \label{T7_NiFe}
\end{figure*}

Until recently, heavier elements such as iron and nickel were only observed in the comae of comets passing very close to the Sun when temperatures are sufficiently high to sublimate refractory material \cite{Jones2018}. However, neutral FeI and NiI emission lines have been found to be very common in the coma of comets as far as 3.25~au from the Sun \cite{Manfroid2021}. These lines are located in the CUBES wavelength range, as shown by the spectrum of the coma of comet C/2016 R2 (PANSTARRS) in Fig.~\ref{T7_NiFe}. The FeI and NiI emission lines were also detected in the coma of the interstellar comet 2I/Borisov \cite{Guzik2021,Opitom2021}. To date, there are no apparent
differences between the NiI/FeI ratio in the comae of solar system comets compared to those of 2I/Borisov, nor between different families of solar system comets \cite{Manfroid2021}. However, the uncertainties on these estimates are quite large and greater near-UV sensitivity is needed to improve the quality of these measurements. 

The detection of NiI and FeI lines in a large number of comets poses very interesting questions. The first is how iron and nickel are released in the coma at the relatively low temperatures prevailing at several au from the Sun. Sulphides, organometallic complexes (e.g. [Fe(PAH)]$^+$), or carbonyls (e.g. Fe(CO)$_5$ and Ni(CO)$_4$) have been suggested as potential reservoirs for the gaseous FeI and NiI \cite{Manfroid2021}. Carbonyls in particular are expected to sublimate at temperatures as low as a few hundred kelvins. However, no consensus on the origin of FeI and NiI has been reached. Chromium is the next most abundant metal in the Sun after nickel. If carbonyls are the main iron and nickel reservoir, simulations that take into account the sublimation rate of Cr(CO)$_6$ indicate that CrI could be as much as 10\,000 times less abundant than FeI \cite{Manfroid2021}. Unsurprisingly given its expected low abundance, CrI has only been detected in the coma of near-Sun comets so far but, with an emission line (357.87~nm) in the CUBES range, this is a promising direction for future observations.

The greater sensitivity of CUBES could also allow us to measure the abundance of NiI and FeI in the comae of comets with smaller uncertainties than with VLT/UVES and for a larger number of comets. This would significantly increase the sample of comets with NiI and FeI abundances, and allow us to search for differences in the NiI/FeI  ratio among comets. Interestingly, a correlation between the NiI/FeI and C$_2$/CN, C$_2$H$_6$/H$_2$O, and NH/CN abundances was recently uncovered, indicating that the NiI/FeI could be related to the formation conditions of comets \cite{Hutsemekers2021}. The subject of metals in the comae of comets is still in its infancy; more questions, beyond the mechanism by which they are released, are raised by the discovery, including what the metals can tell us about the origin and evolution of comets. For example, can these gaseous atomic metal detections be related to the unexpected presence of significant proportions of high-formation-temperature metals/minerals in  samples returned by the Stardust mission, which were used to argue for large scale mixing of material in the solar system's protoplanetary disc \cite{Brownlee2006}? The broader survey of metals in comets possible with a sensitive instrument like CUBES will open a whole new area of research in comet composition, formation, and evolution.


 \section{Other observations}
\label{other}
We have outlined above some of the areas of cometary science where CUBES will provide the most significant improvements. However, a high-resolution, near-UV spectrograph will also enable the observation of other interesting features in the comae of comets. In particular, CUBES covers the CN B-X (0-0) band around 388 nm. The spectral resolution of CUBES is insufficient to measure the isotopic ratios of C and N in the coma, but its sensitivity might permit the detection of CN at record heliocentric distances, helping us to better understand the distant activity of comets. 

Another interesting species present in the coma of comets and that emits in the UV is S$_2$. However, this is very difficult to observe -- it has electronic transitions in the UV and a very short lifetime, meaning that it can only be detected for comets very close to the Earth. Even though it could provide a useful insight into the poorly understood sulphur chemistry of cometary ices, S$_2$ has only been detected in a handful of comets so far \cite{Bockelee2004}. However, it has emission bands in the 305-400 nm region that could be detected with CUBES . 

\section{Conclusions}
\label{sec:conclusions}

Measurements of the D/H and N$_2$/CO ratios, and of composition in general (metals and molecular gasses), all give different constraints on comet formation and evolution. The most comprehensive picture will be built up by combining these pieces of evidence, which can all be derived from the same set of observations with CUBES. The increased sensitivity of CUBES compared with existing instruments will enable a broad survey of comets with consistent measurements, which will be a big step forward for the field. It will also enable measurements, particularly of OD/OH, of a wider diversity of comets than currently possible; rather than just those that are exceptionally active or close to Earth, CUBES will be capable of surveying more typical comets. This will allow us to assess what is really representative of the population, and the ranges in these properties, both within and between dynamical classes from different source regions as well as how much comets may have contributed to the delivery of Earth's water. Pushing the sensitivity of CUBES to its limits will enable gas detections from exceptionally faint comets, including MBCs and comets showing activity at very large distances from the Sun. In turn, these measurements will help quantify how much water is hidden in the asteroid belt and to explain what drives cometary activity in the outer solar system, which is one of the most important questions in the field today.


%
 \section*{Conflict of interest}
 The authors declare that they have no conflict of interest.

\bibliographystyle{spphys}       
\bibliography{Biblio_Comets}   

\end{document}